\documentclass[aps,prc,dlspace,a4paper,twocolumn]{revtex4}
\usepackage{epsfig}
\usepackage{graphics}
\usepackage{color}
\newcommand{\jpsi}{J / \psi}

\newcommand{\old}[1]{}

\newcommand{\be}{\begin{equation}}
\newcommand{\ee}{\end{equation}}
\newcommand{\ba}{\begin{eqnarray}}
\newcommand{\ea}{\end{eqnarray}}
\newcommand{\bi}{\begin{itemize}}
\newcommand{\ei}{\end{itemize}}

\begin{document}
\title{Dissociation of quarkonium in a hot QCD medium: Modification
of the inter-quark potential}
\author{Vineet Agotiya$^{a}$}
\email{vinetdph@iiitr.ernet.in}
\author{Vinod Chandra $^{b}$}
\email{vinodc@iitk.ac.in}
\author{Binoy K. Patra$^{a}$}
\email{binoyfph@iitr.ernet.in}
\affiliation{$^a$Department of Physics, Indian Institute of 
Technology Roorkee, India, 247 667}
\affiliation{
$^{b}$Department of Physics,
Indian Institute of Technology, Kanpur, Kanpur-208 016, India}
\date{\today}
\begin{abstract}
We have studied the dissociation of heavy quarkonium states in a
hot QCD medium by investigating the medium modifications to a heavy quark 
potential. Our model shows that in-medium modification causes
the screening of the charge in contrast to the screening of the 
range of the potential. We have then employed 
the medium-modified  potential to estimate the dissociation pattern of 
the charmonium and bottomonium states and also explore how the 
pattern changes as we go from the perturbative to nonperturbative domain 
in the Debye mass. The results are in good agreement with the other current 
theoretical works both from the spectral function analysis and 
the potential model study.
\end{abstract}
\maketitle


\vspace{1mm}

\section{Introduction}
The study of the fundamental forces among quarks and gluons is an essential
key to the understanding of QCD and the occurrence of different phases which
are expected to show up when going from low to high temperatures  and/or
baryon number densities. For instance, at small or vanishing temperatures 
quarks and gluons get confined inside a hadron by the strong force while 
at high temperatures a quite different medium consisting of quarks 
and gluons known as quark-gluon plasma (QGP) is expected. One of 
the most important features of the QGP formation is the 
color screening of static chromo-electric 
fields~\cite{mclerran}. The suppression of heavy quarkonia ($J/\psi,\,\chi_c,
\,\psi^{'},\,\Upsilon$) due to the color screening analogous to Debye  
screening in QED plasma, has long been proposed as a probe of deconfinement 
in a dense partonic medium. In the deconfined state, the interaction 
between heavy quarks and antiquarks gets reduced due to color screening
leading to a suppression in $J/\psi$ yields~\cite{matsui,dks}.

Thus quarkonia at finite temperature are an important tool to know
the status of the matter (confined/deconfined) formed in heavy ion 
collisions (see, e.g., Ref.~\cite{Sat07}).
Many efforts have been devoted to determine the dissociation temperatures of
$Q\bar{Q}$ states in the deconfined medium, using either lattice calculations
of quarkonium spectral functions
\cite{Asa04,Dat04,Ume05,Iid06} or nonrelativistic
calculations based upon some effective (screened) potentials
\cite{mocsyprd,Dig01,Shu04,Alb05,Won07,Alb07}.
However, the properties of the heavy 
quarkonia states determined from the screened potentials do have a poor 
matching with the results obtained from the lattice spectral functions.
None of the potential model studies and spectral 
functions in lattice to study  quarkonia give a complete framework 
to study the properties of quarkonia at finite temperature. 
It is not yet clear at the moment up to what extent one may understand the 
modifications of quarkonium spectral functions in terms of the
Debye screening picture. One should not expect a precise quantitative 
agreement with the lattice correlators because of the uncertainties 
coming from a variety of sources~\cite{Alb07}. Not only is the determination 
of the effective potential still an open question but also there
are other related issues 
such as relativistic effects, thermal width of the states and 
contribution from quantum corrections that need to be taken care of.
On the other hand, lattice 
correlators are also affected by their own uncertainties. These may be due
to the use of different lattices (isotropic or anisotropic).
Additionally, the finite lattice-spacing might significantly alter the
continuum part of the spectrum. However,
some degree of qualitative agreement had been found for the $S$-wave
correlators. This finding was somehow ambiguous for the
$P$-wave correlators and the temperature dependence of the potential model 
was even qualitatively different from the lattice one.

In a recent work, Umeda~\cite{ume07} found that lattice
calculations of meson correlators at finite temperature
contain a constant contribution due to the presence of
zero modes in the spectral functions. The presence
of a zero mode in the vector channel had already been
discussed in the literature while in the $P$-wave channels it had
generally been overlooked. Recently Alberico {\it et al}~\cite{Alb08} updated
their previous calculation~\cite{Alb07} of quarkonium Euclidean correlators
at finite temperatures in a potential model by including the effect of 
zero modes in the lattice spectral functions. These contributions
cure most of the previously observed discrepancies with lattice 
calculations. This observation supports the use of potential models at finite 
temperature as an important tool to complement lattice studies.

The short and intermediate distance properties
of the heavy quark interaction is important for the understanding
of in-medium modifications of the heavy quark bound states. On the other hand,
the large distance behavior of the heavy quark interaction plays a crucial role 
in understanding the bulk properties of
the QCD plasma phase, {\it viz.} the screening property of the quark gluon
plasma, the equation of state
\cite{Beinlich:1997ia,Karsch:2000ps} and the order parameter (Polyakov loop)
\cite{Kaczmarek:2003ph,Dumitru:2004gd}.

In all of these studies, deviations from perturbative calculations and 
the ideal gas behavior are expected and are indeed found at 
temperatures which are only moderately larger than the deconfinement 
temperature. This calls for  quantitative nonperturbative calculations.
The phase transition in full QCD appears as a crossover
rather than a `true' phase transition with related singularities in
thermodynamic observables (in the high-temperature and 
low density regime)\cite{phaseT}.
Therefore, it is reasonable to assume that the string tension does 
not vanish abruptly above $T_c$.
So one should study its effects on the 
behaviour of quarkonia in a hot QCD medium. 
This issue, usually overlooked in the
literature, is certainly worth investigation.
In the present paper, we considered this potentially
interesting issue by correcting the full Cornell potential with a 
dielectric function embodying the effects of the deconfined medium and not only its Coulomb part as usually done in the literature.
We have found that this leads to a long-range Coulomb potential
with a reduced effective charge 
(inversely proportional to the square of the Debye mass) of the heavy 
quark in addition to the usual Debye-screened form employed in most 
of the literature. With such an effective potential, we investigate the 
effects of different possible choices of the Debye mass on the dissociation 
temperatures of different quarkonium states. Since a Coulomb interaction 
always admits bound states, a criterion has to be adopted to define a
dissociation temperature: a state is then considered to be melted when its
binding energy becomes of the same order as the temperature.
For this  purpose, we consider a gauge-invariant, 
nonperturbative form of the Debye mass 
by Kajantie {\it et al.} \cite{kaj1} and 
study systematically the effects of perturbative and nonperturbative
terms in Debye mass on the dissociation pattern of quarkonia in 
pure gauge, two-flavor, and three-flavor QCD respectively.  Additionally, we consider the lattice parametrized  
form of the Debye mass\cite{mocsyprl}.

The paper is organized as follows.  In-medium 
modifications to heavy quark potential is discussed in Section II.
In subsections IIA and IIB, we study the medium dependence of
quarkonia binding energy and then 
determine their dissociation temperatures 
in a hot QCD medium. Finally, we conclude
in Section III.

\section{In-medium modifications to heavy-quark potential}
Because of the large quark mass $m = m_{c,b} \gg \Lambda_{QCD}$, the
velocity of heavy quarks in the bound state is small and the
binding effects in quarkonia at zero temperature can be understood in
terms of  nonrelativistic potential models \cite{Lucha91}.
More recently, the potential has been derived from QCD using a sequence of
effective field theories (for a review see \cite{nrqcd_rev}).
The present analysis also employs this idea to study the quarkonia
with the nonrelativistic potential model.

Let us now turn our attention to study the
medium modifications to a heavy quark potential which
is considered as the Cornell potential
\begin{equation}
\label{eqc}
 V(r)=-\frac{\alpha}{r}+\sigma r \quad ,
\end{equation}
where $\alpha$ and $\sigma$ are the phenomenological parameters. The former 
accounts for the effective coupling between a heavy quark and its antiquark
and the latter gives the string coupling.

The medium modification enters in the Fourier transform of the 
heavy quark potential as 
\begin{equation}
\label{eq3}
%\tilde {\bf V}(k)=\frac{{\matcal V}(k)}{\epsilon(k)}
\tilde{V}(k)=\frac{V(k)}{\epsilon(k)} \quad ,
\end{equation}
where $\epsilon(k)$ is the dielectric permittivity given in terms 
of the static limit of the longitudinal part of gluon 
self-energy\cite{schneider,weldon}
\begin{eqnarray}
\label{eqn4}
\epsilon(k)=\left(1+\frac{ \Pi_L (0,k,T)}{k^2}\right)\equiv
\left( 1+ \frac{m_D^2}{k^2} \right).
\end{eqnarray} 
Note that the result for the static limit of the dielectric permittivity is the
perturbative one. If one assumes nonperturbative effects
such as the string tension survive even above 
the deconfinement point then the dependence of the dielectric function
on the Debye mass may get modified. 
So there is a {\em caveat} about the
validity of the linear dependence of the dielectric function ($\epsilon$) 
on the square of the Debye mass $M^2_D$. For the sake of simplicity
we put in all the nonperturbative effects 
together in the effective charge ($2\sigma/m_D^2$)
of the medium modified potential.
\begin{figure*}
\vspace{15mm}
\includegraphics[scale=.32]{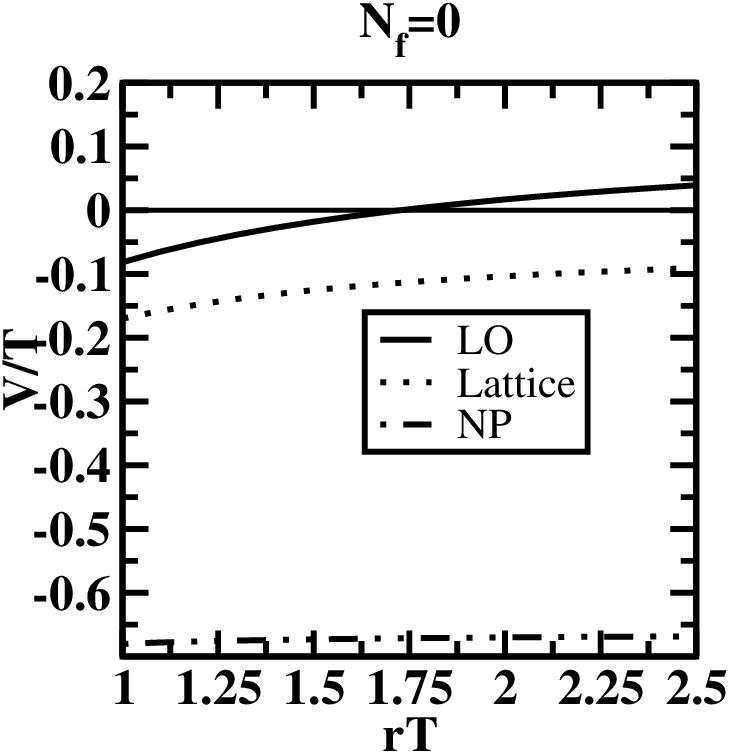} % Here is how to import EPS art
\hspace{5mm}
\includegraphics[scale=.32]{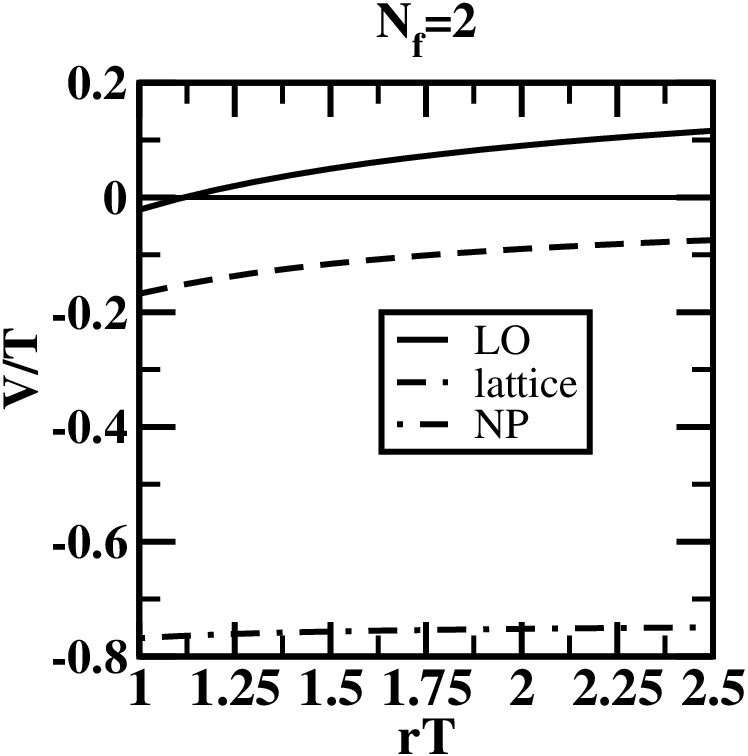}
\hspace{5mm}
\includegraphics[scale=.32]{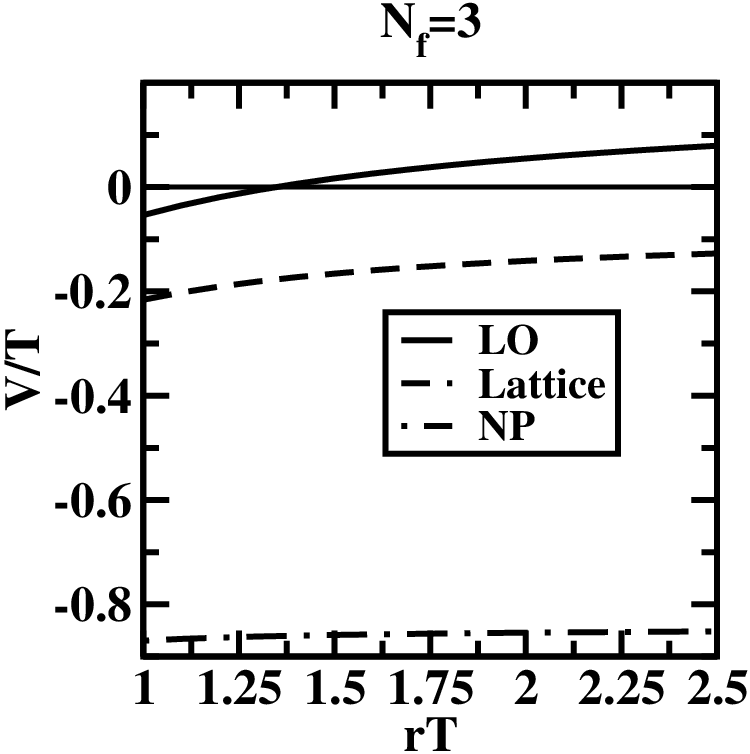}
\caption{The behavior of  $V(r,T)/T$ as a function 
of $r T$ for a fixed $T/T_c=3.32$\cite{pot_1}.
The different curves denote the choice of 
the Debye masses.}
\vspace{25mm}
\end{figure*}

The quantity  $V(k)$ in (\ref{eq3}) is the Fourier transform (FT) of 
the Cornell potential. The evaluation of 
the FT of the Cornell potential is not so straightforward 
and can be done by assuming $r$- as distribution 
($r \rightarrow$ $r \exp(-\gamma r))$. After the evaluation of 
FT we let $\gamma$ tend to zero. Note that the FT of Coulomb 
part is straightforward to compute. The Fourier transform of 
the linear part $\sigma r\exp{(-\gamma r)}$ is 
\begin{eqnarray}
\label{eq-6-3}
=-\frac{i}{k\sqrt{2\pi}}\{\frac{2}{(\gamma-i k)^3}-\frac{2}{(\gamma+ik)^3}\}.
\nonumber
\end{eqnarray}
If we put $\gamma=0$,  we obtain the Fourier transform of $\sigma r$ denoted 
as,
\begin{equation}
\label{eq-6-4}
\tilde{(\sigma r)}=-\frac{4\sigma}{k^4\sqrt{2\pi}}.
\end{equation}

Now the FT of the full Cornell potential can be written as
\begin{equation}
\label{eqn5}
{\bf V}(k)=-\sqrt(2/\pi) \frac{\alpha}{k^2}-\frac{4\sigma}{\sqrt{2}\pi k^4}.
\end{equation}
Substituting Eqs.(\ref{eqn4}) and (\ref{eqn5}) into (\ref{eq3})
and then evaluating its inverse Fourier transform 
one obtains the $r$-dependence of the medium modified 
potential~\cite{chandra1} as:
\begin{eqnarray}
\label{eq4}
{\bf V}(r)&=&(\frac{2\sigma}{m^2_D}-\alpha)\frac{\exp{(-m_Dr)}}{r}\nonumber\\
&-&\frac{2\sigma}{m^2_Dr}+\frac{2\sigma}{m_D}-\alpha m_D
\end{eqnarray}
This potential has a long range Coulombic tail in addition to the 
standard Yukawa term. In the limit $r>>1/m_D$, we can neglect the Yukawa term 
and for large values of temperature the product
$\alpha m_D$ will be much greater than $2\sigma/m_D$. So,
finally the potential (\ref{eq4}) becomes:
\begin{eqnarray}
\label{lrp}
{V(r)}\sim -\frac{2\sigma}{m^2_Dr}-\alpha m_D
\end{eqnarray}
The above form (apart from a constant term) is a Coulombic type as encountered
in hydrogen atom problem with identifying the fine structure constant 
$e^2$ with the effective charge $2 \sigma/m_D^2$. Since
$m_D$ is an increasing function of temperature, the effective
charge $2 \sigma/m_D^2$  gets waned as the temperature is increased 
and finally results in screening of the charge. 
The constant terms in the full potential (\ref{eq4}) 
are introduced by hand in order to remove short-distance medium effects.
However, such terms could arise
naturally from the basic computations of real time static potential 
in hot QCD\cite{laine} and from the real and imaginary time
correlators in a thermal QCD medium\cite{beraudo}.
These terms in the potential
are needed in computing the masses of the quarkonium states and 
to compare the results with the lattice studies.
It is equally important while comparing our effective potential
with the free energy in lattice studies. However, these terms are 
not needed to compare the values of the dissociation temperatures
obtained in our calculation with the values in lattice spectral studies
because we have used different criteria to evaluate the
dissociation temperatures.

It may not be out of context to mention that the expression for the potential
in a hot QCD medium is not the same as the lattice parametrized heavy
quark free-energy in the deconfined phase (which is basically a screened
Coulomb, for the exact form we refer the reader to Refs.\cite{hsatz,shuryak}).
As emphasized by Dixit\cite{dixit} that one-dimensional Fourier
transform of the Cornell potential in the medium yields the similar form
as used in the lattice QCD to study the quarkonium properties 
which assumes the one-dimensional color flux tube structure. 
However, at finite temperature that may not be
the case since the flux tube structure may expand in more 
dimensions\cite{hsatz}.
Therefore, it is better to consider the three-dimensional form of the medium
modified Cornell potential which has been done exactly in the present work.
The medium modified potential thus obtained has a Coulomb tail in
addition to the screened Coulomb part. The strength of the Coulombic
part decreases with the increase in temperature and at a certain 
temperature one may ignore it.
\begin{figure*}
\vspace{15mm}
\includegraphics[scale=.32]{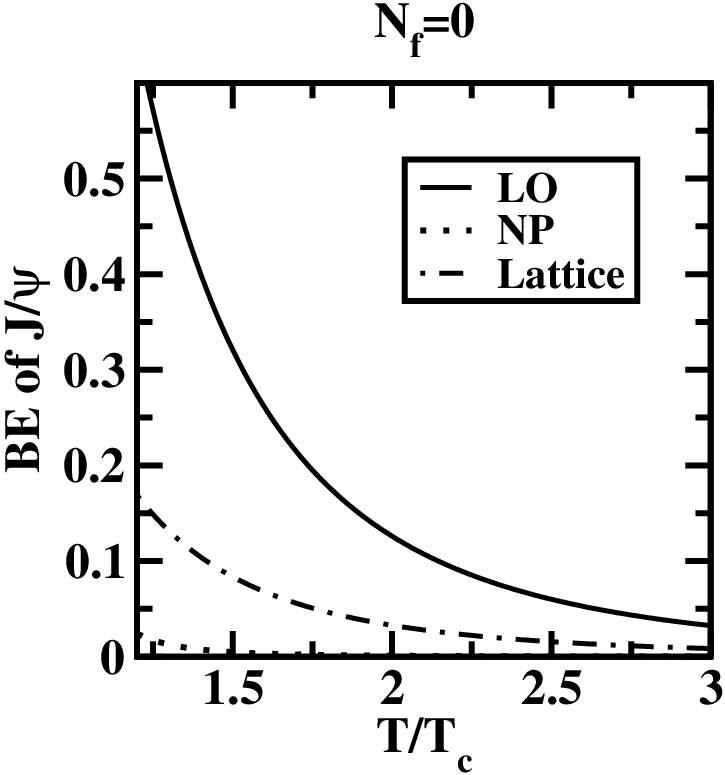} % Here is how to import EPS art
\hspace{5mm}
\includegraphics[scale=.32]{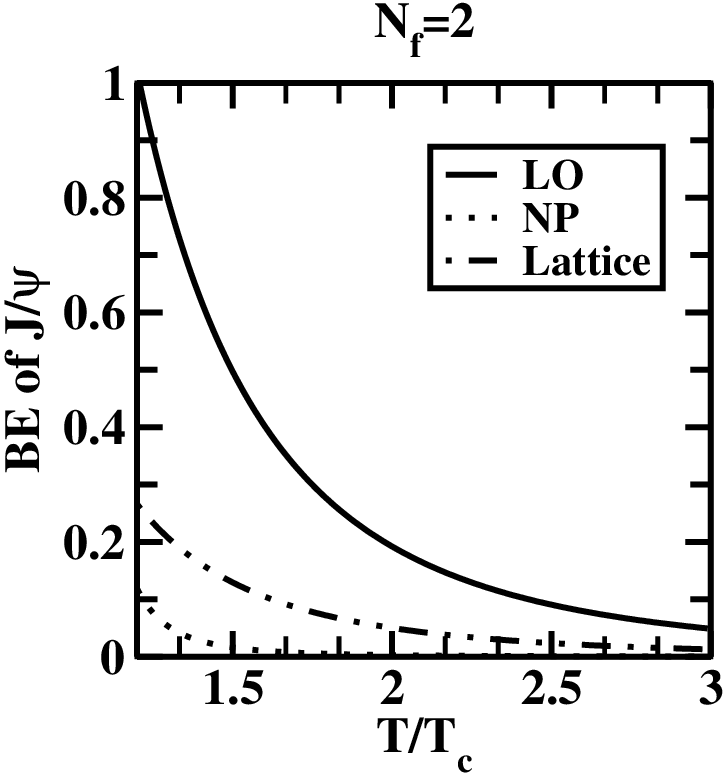}
\hspace{5mm}
\includegraphics[scale=.32]{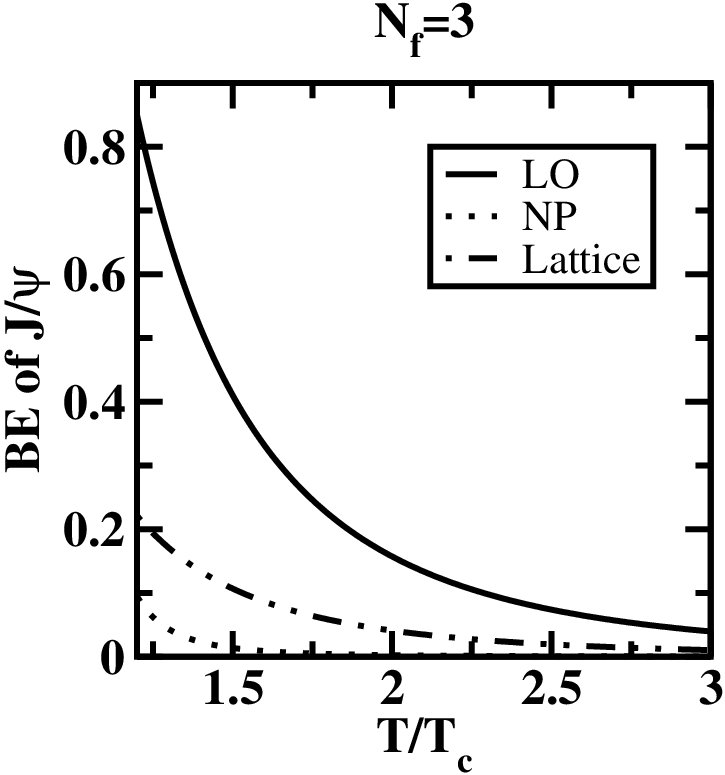}
\caption{The temperature dependence of $J/\psi$ binding energy (in GeV). 
The different curves denote the choice of the
Debye masses.}
\vspace{15mm}
\end{figure*}

To compare our in-medium effective potential 
with the color-singlet free-energy\cite{pot_1} extracted from the
lattice data which do not display at all any long-range term, 
we have plotted the full effective potential 
from (\ref{eq4}) as a function of $r T$ in Fig. 1.
The lattice free energy goes to zero much faster than our effective
potential due to the presence of the Coulomb tail.
However, our potential employing the nonperturbative form of the Debye mass
deviates largely from the lattice results~\cite{pot_1}.

Let us now proceed to study the charmonium and bottomonium spectrum and 
their binding energy with three possible choices of the
Debye mass. Additionally, we take advantage of all
the available lattice data, obtained not only in
quenched QCD ($N_f=0$) but also including two, and more recently, three light
flavors. This enables us to study the flavor dependence of the
dissociation process, a perspective not yet achieved by the parallel studies of
the spectral functions.

\subsection{Binding energy of heavy quarkonia}
Spectral function method defines binding energy  of a quarkonium state 
as the distance between the peak position
and the continuum threshold, $E_{bin}=2 m_{c,b}+V_{\infty}(T) -M~$
with $M$ being the resonance mass. In our case, it is
defined as the `ionization potential' because of the similarity
of our approximated effective potential (\ref{lrp}) with the 
hydrogen atom problem.
Schr\"{o}dinger equation gives the energy eigenvalues for the 
ground states and the first excited states for charmonium ($\jpsi$, $\psi^\prime$ 
etc.) and bottomonium ($\Upsilon$, $\Upsilon^\prime$ etc.) spectra.
Invoking the translational invariance, we can ignore the constant
term in (\ref{lrp}) and the energy of the $n$th eigenstate is given by Bohr's theory:
\begin{eqnarray}
\label{bind1}
E_n=-\frac{E_I}{n^2} \quad; \quad E_I=\frac{m_Q\sigma^2}{m^4_D},
\end{eqnarray}
where $m_Q$ is the mass of the heavy quark and $E_I$ is the energy of the
$Q\bar{Q}$ state in the first Bohr state. The allowed energies 
for $Q\bar{Q}$ states are $E_n=-E_I, -\frac{E_I}{4},-\frac{E_I}{9},
\cdot \cdot \cdot$.
These energies are known as the ionization
potentials/binding energies for the $n$th bound states. 
It becomes a temperature-dependent quantity through
the temperature dependence in Debye mass and it decreases with 
the increase in temperature.

There are other states in the charmonium and bottomonium 
spectroscopy, {\it viz.} $\chi_c$'s and $\chi_b$'s for which the determination 
of the medium-dependent binding energy is beyond the scope of our present 
calculation. For $\chi_c$'s and $\chi_b$'s, one should take into account
the spin dependence of the quark-antiquark potential~\cite{eichten}.
\begin{figure*}
\vspace{15mm}
\includegraphics[scale=.31]{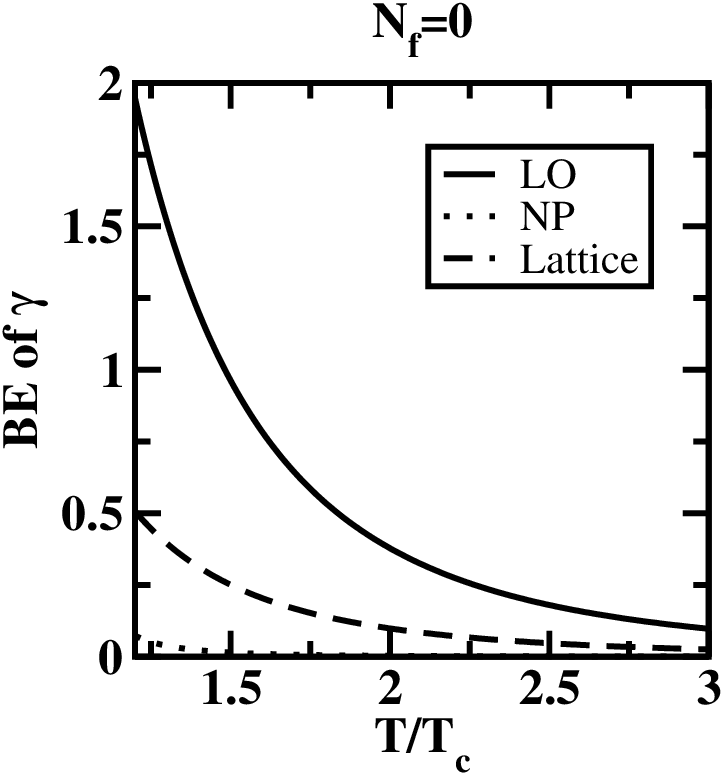} % Here is how to import EPS art
\hspace{5mm}
\includegraphics[scale=.31]{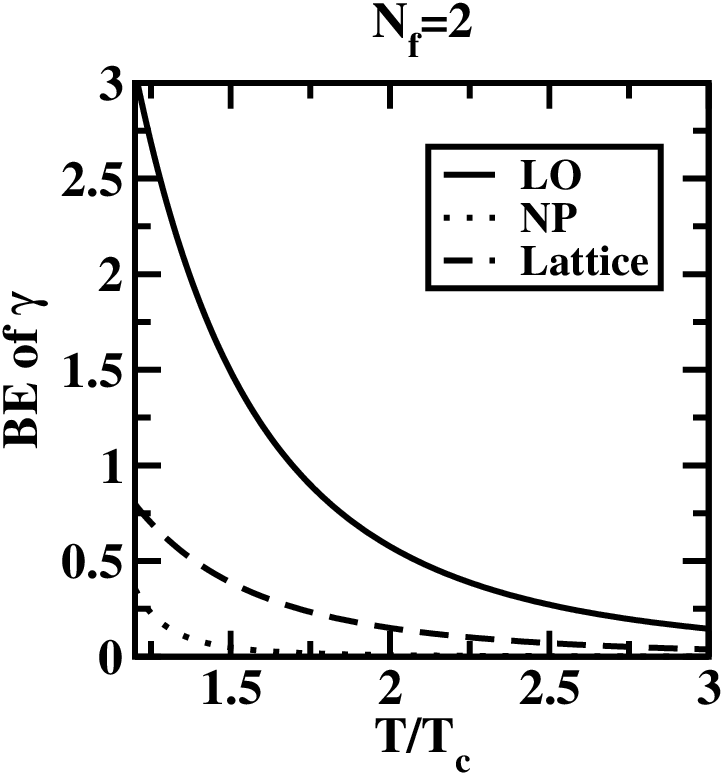}
\hspace{5mm}
\includegraphics[scale=.31]{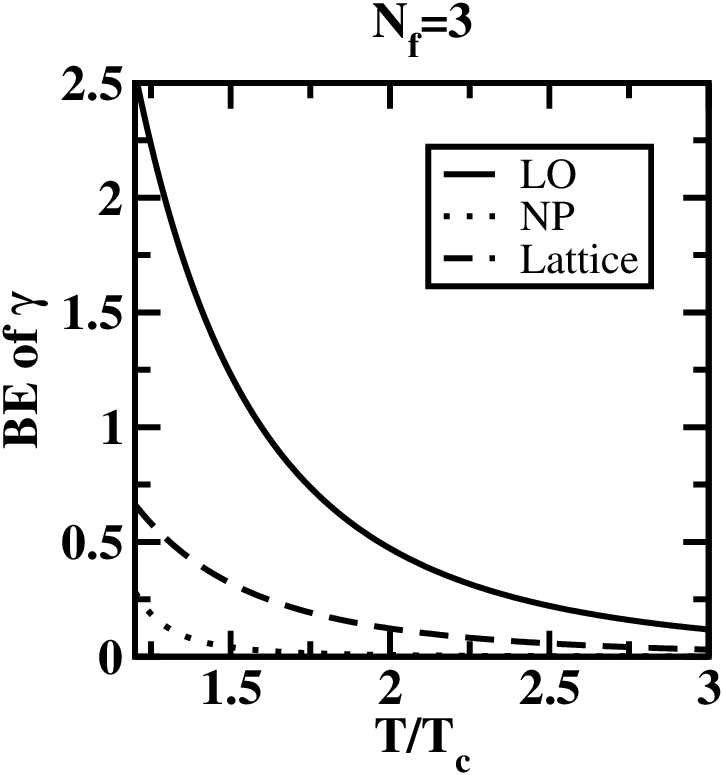}
\caption{Dependence of $\Upsilon$ binding energy(in $GeV$) on 
temperature $T/T_c$.}
\vspace{5mm}
\end{figure*}
Figures 2 and 3 show the variation of binding energy (in GeV) with the 
temperature (in units of critical temperatures) for $J/\psi$ and 
$\Upsilon$, respectively.
Similar variations for other quarkonia ($\psi^\prime$, $\Upsilon^\prime$)
can also be shown.

In the present analysis, we consider three possible forms of the
Debye masses, {\it viz.} the leadingorder term in QCD coupling
($m_D^{\rm{LO}}$), nonperturbative corrections to it
($m_D^{\rm{NP}}$), and the lattice parametrized
form ($m_D^L$) to study the dissociation phenomena of quarkonium
in a hot QCD medium. The Debye mass at high temperature in the
leading-order is known from long time and is
perturbative\cite{shur} in nature. Recently 
Kajantie {\it et al.}~\cite{kaj1} computed the nonperturbative contributions
of $O(g^2T)$ and $O(g^3T)$ from a three-dimensional
effective field theory which we consider in the present work.
However we also consider the Debye mass obtained by fitting the (color-singlet)
free energy in lattice QCD~\cite{mocsyprl}.
Their forms are given below:\vspace{.001in}\\
\ba
\label{notation}
m^{LO}_D&=&g(T) T \sqrt{\frac{N}{3}+\frac{N_f}{6}} \nonumber\\
m^{NP}_D&=& m_D^{\rm{LO}}+{Ng^2T\over4\pi}\ln{ m_D^{\rm{LO}}\over
g^2T}+c_{{}_N} g^2T + d_{N,N_f} g^3 T\nonumber\\
m^{L}_D&=&1.4 m^{LO}_D \quad,
\ea
where the coefficient $c_{{}_N}$ captures the  
nonperturbative effects and 
$d_{N,N_f}$ is related to the choice of the scale in $m_D^{\rm LO}$. We employ
the two-loop expression for QCD coupling constant at finite
temperature\cite{shro} and choose the renormalization scale determined in \cite{shaung}.

Note that different curves in each figure denote the choice of 
the Debye masses (\ref{notation}) used to calculate the binding energy from 
(\ref{bind1}). There is a common observation in Figs. 2,3 that there is
a strong decrease in binding energy with the increase in temperature.
In particular, binding energies obtained from 
$m^{LO}_D$ and $m^L_D$ give realistic variation with the 
temperature. The temperature dependence of the binding energies shows 
a qualitative agreement with the similar variations shown in \cite{mocsyprl}.

However, when we employ the nonperturbative form of the Debye mass $m_D^{NP}$ 
the binding energies become unrealistically
small compared to the binding energy at $T=0$ and also compared to 
the binding energies employing $m^{LO}_D$ and $m^{L}_D$.
This can be understood by the fact that the value of $m^{NP}_D$
is significantly larger than both $m^{LO}_D$ and $m^{L}_D$.
This observation hints that the present 
form of the nonperturbative corrections to the Debye mass 
may not be the complete one, the situation may change 
when nonperturbative corrections of higher order 
$O(g^4T)$ are added to the Debye mass 
and then use it to calculate the binding energy. 

Thus a study of the temperature dependence of the binding energy is poised
to provide a wealth of information about the dissociation
pattern of quarkonium states in thermal medium which will now be 
used to determine the dissociation temperatures of different states.

\subsection{The dissociation temperatures for heavy quarkonia}
Dissociation of a 
two-body bound state in a thermal medium can be understood as:
when the binding energy of a resonance state
drops below the mean thermal energy of a parton, 
the state becomes feebly bound. The thermal
fluctuations then can destroy it by transferring energy and exciting the
quark-antiquark pair into its continuum.
The spectral function technique in potential models defines the dissociation
temperature as the temperature above which the quarkonium spectral
function shows no resonance-like structures but the widths shown 
in spectral functions from current potential model calculations
are not physical. The broadening of states with the increase in 
temperature is not included in any of these models. 
In Ref.\cite{mocsyprl}, the authors
argued that one need not to reach binding energy ($E_{bin}$) to
be zero for the dissociation.
Rather a weaker condition $E_{bin}< T$ causes a state weakly bound and 
the thermal fluctuations can then destroy it. 
Since the (relativistic) thermal energy of the
partons is $3T$, the lower bound on the dissociation temperature ($T_D$)
is obtained from the relation
\begin{equation}
\label{tdiss}
\frac{1}{n^2} \frac{m_Q\sigma^2}{m^4_D(T_D)}= 3 T_D~,
\end{equation}
where string tension ($\sigma$) is taken as 0.184 ${\rm{GeV}^2}$,
and critical 
temperatures ($T_c$) are taken $270 MeV$, $203 MeV$ and $197 MeV$ for pure, 
two-flavor and three-flavor QCD medium, respectively\cite{zantow}.
The dissociation temperatures for 
the ground states and the first excited states of 
$c\bar{c}$ and $b\bar{b}$ are listed in Table I with the Debye mass 
in the leading-order $m^{LO}_D$. 
It is seen from Table I that $\psi^\prime$ is 
dissociated in the vicinity of critical temperature while 
$J/\psi$ and $\Upsilon^\prime$ are dissociated 
around 1.2 $T_c$. $\Upsilon$ is dissociated at a relatively 
higher temperature $1.6T_c$. These values 
agree quantitatively with the recent values reported by Mocsy and 
Petreczky~\cite{mocsyprl}. On the other hand, when we use
Debye mass from lattice parametrized free energy ($m^{L}_D$), the 
values  become much lowered than the leading-order
results (see Table III). However, nonperturbative corrections to the Debye mass
($m_D^{NP}$) make the values unrealistically small. These 
observations can be understood from the hierarchy in their
numerical values: $m^{LO}_D<m^{L}_D<m^{NP}_D$.

\begin{table}
\label{table1}
\caption{Lower (upper) bound on the dissociation temperature($T_D$) 
for the quarkonia states (in units of $T_c$) using the leading-order
term in the Debye mass $m^{LO}_D$.}
\centering
\begin{tabular}{|l|l|l|l|l|}
\hline
State &Pure QCD & $N_f=2$&$N_f=3$\\
\hline\hline
$\jpsi$&1.1 (1.4) &1.3 (1.7) &1.2 (1.6) \\
\hline
$\psi'$&0.8 (1.0) &0.9 (1.2) &0.9 (1.1) \\
\hline
$\Upsilon$&1.4 (1.8) &1.7 (2.1) &1.6 (2.0) \\
\hline
 $\Upsilon'$&1.0 (1.3) &1.2 (1.6) &1.2 (1.5) \\
\hline
\end{tabular}
\end{table}
%%%%%%%%%%%%%%%%%%%%%%%%%%%%%%%%%%%%%%%%%%%%%%%%%%%%%%%%%%%%%5
\begin{table}
\label{table2}
\caption{same as Table I but using full potential (\ref{eq4}).}
\centering
\begin{tabular}{|l|l|l|l|l|}
\hline
State &Pure QCD & $N_f=2$&$N_f=3$\\
\hline\hline
$\jpsi$&1.2 (1.5) &1.3 (1.72) &1.3 (1.7) \\
\hline
$\psi'$&0.8 (1.2) &1.0 (1.4) &1.1 (1.2) \\
\hline
$\Upsilon$&1.4 (1.8) &1.7 (2.3) &1.6 (2.1) \\
\hline
 $\Upsilon'$&1.0 (1.4) &1.2 (1.7) &1.2 (1.5) \\
\hline
\end{tabular}
\end{table}

\begin{table}
\label{table3}
\caption{Lower (upper) bound on the dissociation temperatures
using the lattice parametrized form of the Debye mass $m^{L}_D$.}
\centering
\begin{tabular}{|l|l|l|l|l|}
\hline
State &Pure QCD & $N_f=2$&$N_f=3$\\
\hline\hline
$\jpsi$&0.8 (1.0) &0.9 (1.2) &0.9 (1.1) \\
\hline
$\psi'$&0.5 (0.7) &0.7 (0.8) &0.6 (0.8) \\
\hline
$\Upsilon$&1.0 (1.3) &1.2 (1.6) &1.2 (1.5) \\
\hline
 $\Upsilon'$&0.7 (0.9) &0.9 (1.1) &0.8 (1.0) \\
\hline
\end{tabular}
\end{table}

The fact that  $m^{NP}_D$
leads to unrealistic smaller values of dissociation
temperatures does not imply that
one should ignore the nonperturbative terms in the Debye mass.
In fact, nonperturbative terms cannot be ignored in the regime where
coupling is strong which is indeed the case dealt with. It would rather 
be of interest to
raise the question why this nonperturbative result obtained with a
dimensional reduction is not in agreement with the Debye mass
arising from Polyakov-loop correlators.
This could be partially due to the arbitrariness in 
the definition of dissociation temperature, since strictly 
speaking a Coulomb potential always admits bound states in its spectrum. 
Indeed the choice of the average thermal energy $3T$ is not rigid 
because even at low temperatures
$T < T_c$ (say) the Bose/Fermi distributions of partons will
have a high energy tail with partons of mechanical energy
greater than the binding energy. So, we calculate the upper bound of the 
dissociation temperatures by 
replacing the average thermal energy $\sim T$ which is listed within the
first bracket in the tables where the values are increased by 30\% 
approximately.

The results for the dissociation temperatures of various quarkonia
listed above in Table I and Table III are obtained by dropping all 
the finite-range terms in the full effective potential (\ref{eq4}). 
As mentioned earlier, for the $s$-wave states 
this  leads to an analytically solvable Coulomb potential.
To see the effects of the finite-range terms in (\ref{eq4}), we 
solve the Schr\"odinger equation numerically with the full effective 
potential(\ref{eq4})
and determine the energy spectrum of the ground and the first excited states of the
charmonium and bottomonium spectrum with the Debye mass in the 
leading-order.
We find that the dissociation temperatures change by  $\sim$ 10 20\% shown
in  Table II. The dissociation temperature 
for $J/\Psi$ in the pure gauge case becomes $1.2 T_c$
which was earlier $1.1T_c$ (Table I) and for $\psi^\prime$, it now becomes 
$1.0T_c$ for $N_f=2$ which was earlier $0.9 T_c$. The same trend follows for other 
charmonium and bottomonium states. This slight increase in the dissociation
temperatures is caused by the increase in the binding energies
due to the finite-range terms in the full potential (\ref{eq4}).

Finally, to compare our results with a recent calculation~\cite{mocsyprl} 
having the same input based on a potential study for a three-flavor QCD with $T_c=192 MeV$, we calculated the upper bound of dissociation temperatures 
in Table IV and V with the same form of 
Debye mass used in Ref.\cite{mocsyprl}.
It shows a good agreement with their results~\cite{mocsyprl}. 
However, the agreement holds well even with the full Cornell potential.

\begin{table}
\caption{Upper bound on the dissociation temperatures ($T_D$) 
with $T_c=192$ MeV \cite{mocsyprl} using the lattice parametrized form of the
screening mass($m^{L}_D$).}
\centering
\begin{tabular}{|c|c|c|c|c|c|c|}
\hline
 State & $\psi^\prime$ & $\jpsi$ & $\Upsilon^\prime$ & 
& $\Upsilon$ \\
\hline
$T_D$ & $\le 0.9T_c$ & 1.2$T_c$ & 1.1$T_c$ & & 1.6$T_c$ \\
\hline
\end{tabular}
\end{table}

\begin{table}
\caption{Same as Table IV but using the full potential (\ref{eq4}).}
\centering
\begin{tabular}{|c|c|c|c|c|c|c|}
\hline
 State & $\psi^\prime$ & $\jpsi$ & $\Upsilon^\prime$ &
& $\Upsilon$ \\
\hline
$T_D$ & $\le 1.0T_c$ & 1.3$T_c$ & 1.2$T_c$ & & 1.6$T_c$ \\
\hline
\end{tabular}
\end{table}

\section{Conclusions and Outlook}
In conclusion, we have studied the dissociation phenomena of quarkonia 
in a hot QCD medium by investigating the in-medium modifications to a heavy 
quark potential. 
We have found that medium modification causes a dynamical screening
of color charge which in turn, leads to a temperature dependent 
binding energy. We have systematically studied the 
temperature dependence of binding energy for the 
ground and first excited states of 
charmonium and bottomonium spectra in pure and realistic QCD medium. 
We have then determined the dissociation temperatures employing 
the perturbative result of Debye mass ($m_D^{\rm{LO}}$) and the lattice 
parametrized form $m_D^L$. Our estimates are consistent with
the finding of recent theoretical works based on 
potential models~\cite{mocsyprl}. 
However, these values are significantly smaller than the 
predictions of Refs.\cite{Sat07,Dat04,Alb07,gert} based on 
the first principle lattice calculations which are however plagued
by its inherent uncertainties.
In contrast, the inclusion of 
nonperturbative  contributions to Debye mass lowers the dissociation 
temperatures substantially which looks unfeasible. Thus, this study 
provides us a handle to decipher the extent up to which and how much
nonperturbative effects should be incorporated into the Debye mass.

In brief, $\jpsi$ is found to be dissociated at temperature above the critical
temperature (around $1.2T_c$) when the leadingorder term in
the Debye mass has been employed. However,  it is dissociated
just below the $T_c$ when lattice 
parametrized form of Debye mass (nonperturbative) has been employed. 
This finding ensues a basic question about the nature of dissociation 
of quarkonium in a hot QCD medium.

Finally, our approach based on the in-medium modifications
provides charmonium and bottomonium
dissociation temperatures which agree nicely with  recent quarkonium
spectral function studies using a potential model~\cite{mocsyprl}. 
This is true only for the perturbative 
result for the Debye mass but nonperturbative corrections to it
make the melting temperatures too low to compare to the spectral
analysis of the lattice temporal correlator of the mesonic current.
This leaves an open problem of the agreement between these two kinds 
of approaches.
This could partially be due to the arbitrariness in the definition of
dissociation temperature.
To examine this point we have estimated the upper bound of the 
dissociation temperatures from the condition: $E_{bin}=T$. We found 
that these estimates 
obtained by employing the lattice parametrized Debye mass 
show good agreement with the predictions in \cite{mocsyprl} which
was not true for the earlier definition 
(\ref{tdiss}) : $E_{bin}=3T$.
However, a numerical solution of the Schr\"odinger equation with the
full effective potential (\ref{eq4}) gives in general a slightly 
higher value of the dissociation temperatures.
For $\chi_c(\chi_b)$ melting temperatures, one would start with the 
spin-dependent heavy quark potential and follow the same  procedure.
We will look into it in the future.
Finally it would be of interest to study the corresponding quarkonium 
spectral function (and temporal
correlator) after giving these states a thermal width.

\noindent {\bf Acknowledgments:}
We acknowledge  V J Menon, A. Ranjan and V. Ravishankar for fruitful 
discussions and CSIR, New Delhi, India for the financial support.

\end{document}